\documentclass[a4paper,11pt]{article}
\pdfoutput=1 
\usepackage{jheppub} 

\usepackage[T1]{fontenc} 

\pdfoutput=1
\usepackage{amsmath,amssymb}
\usepackage{amsfonts}
\usepackage{bm,bbm}
\usepackage{graphicx}
\usepackage{mathrsfs}
\usepackage{slashed}
\usepackage{booktabs}
\usepackage{tabu}
\usepackage[dvipsnames]{xcolor}
\usepackage[normalem]{ulem}
\usepackage{tikz}
\usepackage{soul}

\usepackage{hyperref}
\hypersetup{colorlinks,citecolor= blue,linkcolor= blue, urlcolor=blue}

\usepackage{xcolor}
\usepackage{ulem}
\usepackage{array}
\usepackage{verbatim}
\usepackage{epsfig}
\usepackage{multirow}

\usepackage{ulem,fancyvrb}
\usepackage{xcolor} 
\newcommand{\zb}[1]{{\color{blue} {#1}}}

\title{Refined Low-Energy Supernova Constraints 
on Lepton Flavor Violating Axions}

\author[a]{Zi-Miao Huang,}
\author[a]{Changqian Li,}
\author[a]{Zuowei Liu}

\affiliation[a]{Department of Physics, Nanjing University, Nanjing 210093, China}

\emailAdd{zimiaohuang@smail.nju.edu.cn}
\emailAdd{changqianli@smail.nju.edu.cn}
\emailAdd{zuoweiliu@nju.edu.cn}

\abstract{The supernova (SN) core, characterized by its extreme temperature and density, serves as a unique laboratory for new-physics searches. Low-energy supernovae (LESNe) provide particularly powerful probes, as their low explosion energies place stringent limits on any additional energy deposition in the mantle by new particles. We present refined LESN constraints on lepton-flavor-violating (LFV) axions and axion-like particles (ALPs) with electron-muon couplings. We consider four production channels in the SN: muon decay, lepton bremsstrahlung, electron-muon coalescence, and semi-Compton scattering, the last of which is investigated here for the first time in the context of LFV-ALPs. We find that muon decay dominates in the low-mass regime, electron–muon coalescence in the high-mass regime, and semi-Compton scattering in the intermediate-mass range.  To derive accurate limits, we compute both the energy transfer from the SN core to the mantle and the energy loss due to ALP production in the mantle, which can be substantial for both large and small couplings---the latter case, to our knowledge, not previously noted in the literature. We find that LESNe provide the most stringent constraints on the parameter space for ALP masses above $\sim 110$ MeV.  These refined results strengthen previous SN bounds and highlight the exceptional sensitivity of LESNe to LFV new physics.}

\begin{document}

\maketitle
\flushbottom

\section{Introduction}
\label{sec:intro}

Axions are hypothetical pseudoscalar 
bosons that arise
in extensions of the Standard Model, 
originally proposed to solve the strong CP problem
\cite{Peccei:1977hh,Wilczek:1977pj,Weinberg:1977ma,Peccei:2006as}.
Besides the original axions,
a wide class of pseudoscalar particles--collectively referred to as 
axion-like particles (ALPs)--naturally emerge in numerous well-motivated 
new physics models beyond the Standard Model
\cite{Gelmini:1980re,Davidson:1981zd,Wilczek:1982rv,Svrcek:2006yi}.
While most ALP searches focus on flavor‐conserving couplings, 
lepton-flavor-violating (LFV) interactions can naturally 
arise from various ultraviolet completions or radiative corrections
\cite{Davidson:1981zd,Wilczek:1982rv,Anselm:1985bp,Feng:1997tn,Bauer:2016rxs,Ema:2016ops,Calibbi:2016hwq,Choi:2017gpf,Chala:2020wvs,Bauer:2020jbp}.
At the GeV scale and above,
colliders and fixed‐target facilities 
have placed stringent bounds on LFV-ALPs
\cite{Endo:2020mev,Iguro:2020rby,Davoudiasl:2021haa,Cheung:2021mol,Davoudiasl:2021mjy,Araki:2022xqp,Calibbi:2022izs,Batell:2024cdl,Calibbi:2024rcm}.  
In the sub-GeV regime, 
muon decay experiments
\cite{Derenzo:1969za,Jodidio:1986mz,Bryman:1986wn,Bilger:1998rp,TWIST:2014ymv,Bauer:2019gfk,Cornella:2019uxs,PIENU:2020loi,Calibbi:2020jvd,Bauer:2021mvw,Knapen:2024fvh,Jho:2022snj,Knapen:2023zgi},
and supernova (SN) observations
\cite{Calibbi:2020jvd,Zhang:2023vva,Li:2025beu,Huang:2025rmy} 
provide leading constraints on ALPs 
that have an LFV coupling to electrons and muons.

In this work, 
we study constraints on LFV-ALPs from 
low‐energy supernovae (LESNe), a
class of underluminous Type II‐P explosions 
with ejecta energies as low as 
$0.1\ \mathrm{B}=10^{50}\ \mathrm{erg}$, i.e., roughly 
10-100 times dimmer than typical core-collapse SNe 
\cite{Caputo:2022mah}.
LESNe have been identified both observationally
\cite{Chugai:1999en,Pastorello:2003tc,Pastorello:2009pt,10.1093/mnras/stu156,Pejcha:2015pca,Yang:2015ooa,Pumo:2016nsy,Murphy:2019eyu,Burrows:2020qrp,Yang:2021fka,Teja:2024cht}
and in simulations 
\cite{Kitaura:2005bt,Fischer:2009af,Melson:2015tia,Radice:2017ykv,Lisakov:2017uue,Muller:2018utr,Burrows:2019rtd,Stockinger:2020hse,Zha:2021bev}. 
Unlike the conventional SN cooling bounds, which 
require that exotic energy losses not exceed the neutrino luminosity 
\cite{Raffelt:1996wa}, 
LESN limits constrain the total energy deposited 
by new particles in the mantle to remain below the explosion energy of
0.1~B 
\cite{Falk:1978kf,Sung:2019xie,Caputo:2022mah}. 
Recent analyses have shown that LESNe 
can provide powerful probes of new light particles
\cite{Caputo:2022mah, Lella:2024dmx, Alda:2024cxn,
	Chauhan:2023sci, Chauhan:2024nfa, Carenza:2023old,
	Fiorillo:2024upk,Li:2024pcp,Fiorillo:2025yzf,Huang:2025rmy,Fiorillo:2025sln}.

Previous works have explored SN constraints on LFV-ALPs, 
including both the SN cooling limits 
\cite{Calibbi:2020jvd,Zhang:2023vva,Li:2025beu} 
and the LESN limits \cite{Huang:2025rmy}.
However, Ref.~\cite{Huang:2025rmy} focused exclusively on  
LFV-ALPs with masses $m_a > m_\mu + m_e$, 
where only the electron-muon coalescence channel 
and its inverse process (ALP decay) were considered for 
production and absorption, respectively.
In this paper, we extend the analysis 
to the full parameter space, 
including the regime $m_a < m_\mu + m_e$, 
where we incorporate four production processes: 
(1) muon decay, 
(2) lepton bremsstrahlung,  
(3) electron-muon coalescence, and 
(4) semi-Compton scattering. 
The semi-Compton scattering was recently found to 
contribute significantly to axions that couple to 
electrons \cite{Fiorillo:2025sln}.

It has been recently pointed out 
that for strong couplings, energy outflows from the 
mantle to both the core and the exterior of the progenitor  
can be substantial \cite{Fiorillo:2025yzf}. 
We include these effects to obtain 
more reliable LESN constraints in the large-coupling regime. 
For LFV-ALPs, we find that the energy drain from the mantle due 
to ALP production within it can be substantial not only in  
the strong coupling regime 
but also in the weak coupling regime---a feature that, to our knowledge, 
has not been recognized in previous studies.
Moreover, a proper treatment of ALP absorption is crucial  
for deriving accurate LESN constraints. 
In Ref.~\cite{Huang:2025rmy}, 
the survival probability along an ALP trajectory was approximated by 
evaluating the local absorption rate at the production point. 
Here, we explicitly integrate the absorption rate  
along the full ALP propagation path, leading to a more 
accurate evaluation of survival probability. 
Finally, while Ref.~\cite{Huang:2025rmy} used the 
Garching SN profiles \cite{Bollig:2020xdr,garching-profile} 
at one second post-bounce, we employ the Garching SN profiles 
spanning the full 10-second post-bounce evolution, 
thus providing a more robust bound.

The rest of the paper is organized as follows. 
In Section~\ref{sec:model} we introduce the LFV-ALP model. 
In 
Section~\ref{sec:Net:energy:deposition} 
we compute the energy deposition in the SN mantle, 
including both the positive contributions from ALPs produced in the 
SN core and the negative contributions from ALPs produced in the mantle. 
Section \ref{sec:profile} 
provides a brief introduction to the Garching group's 
muonic SN model SFHo-18.8. 
We compute the ALP production rate in Section \ref{sec:production}, 
where we consider four different production channels: 
(1) muon decay, 
(2) lepton bremsstrahlung,  
(3) electron-muon coalescence, and 
(4) semi-Compton scattering. 
We then compute the ALP absorption rate in Section \ref{sec:absorption}. 
We present our results in Section~\ref{sec:results}, and 
summarize our main findings in Section~\ref{sec:summary}.
Finally, Appendix~\ref{sec:ALP:p:semi-compton} contains 
the detailed derivation of 
the Mandelstam variables for the semi-Compton scattering.

\section{Model}
\label{sec:model}

We consider the LFV-ALP model which has the following interaction 
with electrons and muons: \cite{Huang:2025rmy}
\begin{equation}
    \mathcal{L}_{\rm int} = \frac{g_{ae\mu}}{m_e^0+m_\mu}
    \bar e \gamma^\lambda \gamma_5 \mu 
    \partial_\lambda a + {\rm h.c.},
    \label{eq:lagrangian}
\end{equation}
where 
$a$ denotes the ALP field, 
$e$ and $\mu$ are the electron and muon fields, respectively, 
$m_e^0 = 0.511~\mathrm{MeV}$ is the electron mass in vacuum, 
$m_\mu = 105.6~\mathrm{MeV}$ is the muon mass, and 
$g_{ae\mu}$ is the dimensionless coupling constant. 
Note that the interaction Lagrangian in 
Eq.~\eqref{eq:lagrangian} is equivalent to the 
following interaction Lagrangian 
\begin{equation}
    \mathcal{L}_{\rm int} = -ig_{ae\mu} a 
    \bar e \gamma_5 \mu + {\rm h.c.},
    \label{eq:eff-lagrangian}
\end{equation}
when the fermions are on-shell 
\cite{Raffelt:1987yt,Lucente:2021hbp,Ferreira:2022xlw,Li:2025beu}. 
We use the interaction Lagrangian in Eq.~\eqref{eq:eff-lagrangian} in analysis.

\section{Energy deposition in the mantle}
\label{sec:Net:energy:deposition}

ALPs can be copiously produced in the SN core 
owing to the extreme temperature and density conditions. 
As they propagate outward, a fraction of 
these ALPs are reabsorbed in the surrounding mantle, 
thereby transferring energy from the core to the mantle, 
which contributes to the SN explosion energy. 
On the other hand, ALPs produced in the mantle can transfer energy 
back to the core or outside the progenitor star, 
resulting in an energy drain from the mantle, 
which is particularly significant for strong couplings \cite{Fiorillo:2025yzf}.  
Thus, the net energy deposition 
in the SN mantle is given by \cite{Fiorillo:2025yzf}
\begin{equation}
    E_d = E_c - E_m,
    \label{eq:energy-deposition-total}
\end{equation}
where $E_c$ denotes the energy transferred 
to the mantle by ALPs produced in the core, 
and $E_m$ represents 
the energy loss from the mantle due to ALPs 
generated within it.

\subsection{ALP production in the core}
\label{sec:ALP:production:core}

We first compute the energy transferred to the mantle 
by ALPs originating from the core. 
The net energy deposition in the mantle due to these core-produced ALPs 
is defined as the difference between the energy escaping 
the core surface and that escaping the progenitor's outer boundary
\cite{Caputo:2022rca, Caputo:2022mah, Li:2025beu}: 
\begin{equation}
E_c
 = \int dt \int_0^{R_c} d r \int_{m_a^{\prime}(r)}^{\infty} 
 d E_a \frac{d L_a\left(r, E_a, t\right)}{d r \, d E_a}  
\left[ 
\left\langle e^{-\tau_a(R_c,E_a,r)}
\right\rangle
-
\left\langle e^{-\tau_a(R_*,E_a,r)}
\right\rangle
\right],
\label{eq:energy-deposition}
\end{equation}
where 
$t$ is time, 
$r$ is the radial coordinate of ALP production, 
$E_a$ is the ALP energy, 
$\tau_a$ is the optical depth, 
$L_a(r, E_a, t)$ is the ALP luminosity, 
$m_a^\prime(r) = m_a / \mathrm{lapse}(r)$
with 
$m_a$ being the ALP mass 
and $\mathrm{lapse}(r)$ being the gravitational lapse function 
\cite{Caputo:2022mah}, 
and 
$R_c$ and $R_*$ are the radius of the core and the progenitor star, 
respectively. 
The two terms inside the brackets in 
Eq.~\eqref{eq:energy-deposition} describe 
the ALP survival probability (due to reabsorption effects) 
at $R_c$ and $R_*$, respectively. 
In our analysis we use 
$R_c = 20 \,\mathrm{km}$ and 
$R_\ast=5\times10^{8}$ km, 
which corresponds to the progenitor radius 
for the Garching group's muonic model SFHo-18.8 \cite{Caputo:2022mah}.

\begin{figure}
    \centering
    \includegraphics[width=0.25\linewidth]{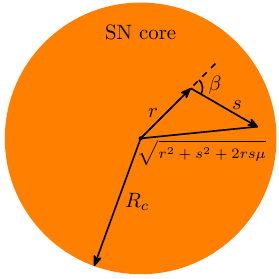}
    \caption{Schematic plot of the production position $r$, the propagation distance $s$, and the angle $\beta$ inside the SN core.}
    \label{fig:DiagrammaticPlot}
\end{figure}

The ALP optical depth is given by 
(see e.g., Refs.~\cite{Caputo:2021rux,Caputo:2022rca,Fiorillo:2025yzf})
\begin{equation}
\tau_a(R,E_a,r,\mu) = 
\int_0^{s_{\rm max}} \frac{ds}{v} \, \Gamma 
\left(E_a,\sqrt{r^2 + s^2 + 2rs\mu}\right), 
\end{equation}
where $v$ is the ALP velocity, 
$s$ is the propagation distance, 
$\Gamma=\Gamma_A-\Gamma_E$ is the reduced ALP absorption rate
with $\Gamma_A$ ($\Gamma_E$) being the absorption (emission) rate, 
and 
$s_{\rm max}= \sqrt{R^2-r^2(1-\mu^2)}-r \mu$, 
where 
$\mu \equiv \cos\beta$ 
with $\beta$ being the angle between the ALP trajectory and the outward radial direction; 
see Fig.~\ref{fig:DiagrammaticPlot} for a schematic plot
of the production position $r$, the propagation distance $s$, and the angle $\beta$
inside the SN core. 
Note that $\sqrt{r^2+s^2+2rs\mu}$ denotes the 
radial coordinate along the trajectory.
Assuming isotropic propagation, 
one obtains the averaged survival probability 
\cite{Caputo:2021rux,Caputo:2022rca,Li:2025beu}
\begin{equation}
\left\langle e^{-\tau_a(R,E_a,r)}
\right\rangle
= \int_{-1}^{1} \frac{d\mu}{2} e^{-\tau_a(R,E_a,r,\mu)}. 
\label{eq:abs-full}
\end{equation}

The ALP luminosity per unit 
radial distance per unit energy is given by 
\footnote{Note that the dimension of luminosity is the same as $dE/dt$, 
and $4\pi r^2 dr$ is the volume of the spherical shell between $r$ and $r+dr$.}
\begin{equation}
    \frac{d L_a\left(r, E_a, t\right)}{d r d E_a }= 4 \pi r^2 \, \mathrm{lapse}(r)^2 (1+2 v_r) E_a \frac{d^2 n_a}{dt \, dE_a}, \label{eq:luminosity}
\end{equation}
where $d^2 n_a/dtdE_a$ is the ALP production rate per unit volume per unit energy, 
and $v_r$ is the radial velocity of the medium.  
Because $v_r \ll 1$ \cite{Caputo:2022mah}, 
in our analysis we neglect $v_r$ in Eq.~\eqref{eq:luminosity}. 
The emission rate is related to the production rate via
\begin{equation}
\Gamma_E = \frac{2\pi^2}{|\textbf{p}_a| E_a} \frac{d^2 n_a}{dt dE_a},     
\end{equation}
where $|\textbf{p}_a| = \sqrt{E_a^2 - m_a^2}$.

\subsection{ALP production in the mantle}
\label{sec:ALP:production:mantle}

We next compute the energy drain by ALPs 
produced in the mantle.
ALPs generated in the mantle and subsequently absorbed in the core 
or escaping the progenitor star lead to an energy loss
from the mantle \cite{Fiorillo:2025yzf}: 
\begin{equation}
E_m=E_m^{(1)}+E_m^{(2)}, \label{eq:mantle-contribution}
\end{equation}
where $E_m^{(1)}$ and $E_m^{(2)}$ 
denote the energy transferred from the mantle 
to the core
and to
the exterior of the progenitor, respectively. 
Following Ref.~\cite{Fiorillo:2025yzf}, 
we compute $E_m^{(1)}$ as follows: 
\begin{align}
E_m^{(1)}
 =& \int dt \int_{R_c}^{R_{*}} d r \int_{m_a^{\prime}(r)}^{\infty} d E_a 
 \frac{d L_a\left(r, E_a, t\right)}{d r \, d E_a} \nonumber \\
&\times \int_{-1}^{-\sqrt{1-R_c^2/r^2}} \frac{d \mu}{2} 
e^{-\tau_a'(E_a,r,\mu)}
\left[1-e^{-\tau_a''(E_a,r,\mu)}\right], 
\label{eq:mantle-to-core}
\end{align}
where 
\begin{equation}
\tau_a'(E_a,r,\mu)
=
\int_0^{s_1} \frac{ds}{v} \, \Gamma
\left(E_a,\sqrt{r^2 + s^2 + 2rs\mu}\right)
\label{eq:tau1}
\end{equation}
is the optical depth between the ALP 
production point and the entry point into the core, and 
\begin{equation}
\tau_a''(E_a,r,\mu) 
= 
\int_{s_1}^{s_2} \frac{ds}{v} \, \Gamma_{\rm}
\left(E_a,\sqrt{r^2 + s^2 + 2rs\mu}\right)
\label{eq:tau2}
\end{equation}
is the optical depth within the core.
The two integration limits in 
Eqs.~(\ref{eq:tau1})and (\ref{eq:tau2}) are given by 
$s_{1,2}=-r\mu\mp[R_c^2-r^2(1-\mu^2)]^{1/2}$, 
which correspond to the entry and exit points of the core.
We compute $E_m^{(2)}$ as follows: 
\begin{equation}
E_m^{(2)}
 = \int dt \int_{R_c}^{R_{*}} d r \int_{m_a^{\prime}(r)}^{\infty} d E_a 
 \frac{d L_a\left(r, E_a, t\right)}{d r \, d E_a}  
\left\langle e^{-\tau_a(R_*,E_a,r)}
\right\rangle.
\label{eq:mantle-to-progenitor}
\end{equation}

Note that for 
$m_a>m_\mu+m_e^0$ and sufficiently large couplings,
ALPs decay mostly within the progenitor 
via $a\rightarrow e^\mp\mu^\pm$,  
resulting in the approximation 
$E_m^{(2)} \approx 0$.

\section{SN profiles}
\label{sec:profile}

\begin{figure}[t]
\centering
\includegraphics[width=0.48\textwidth]{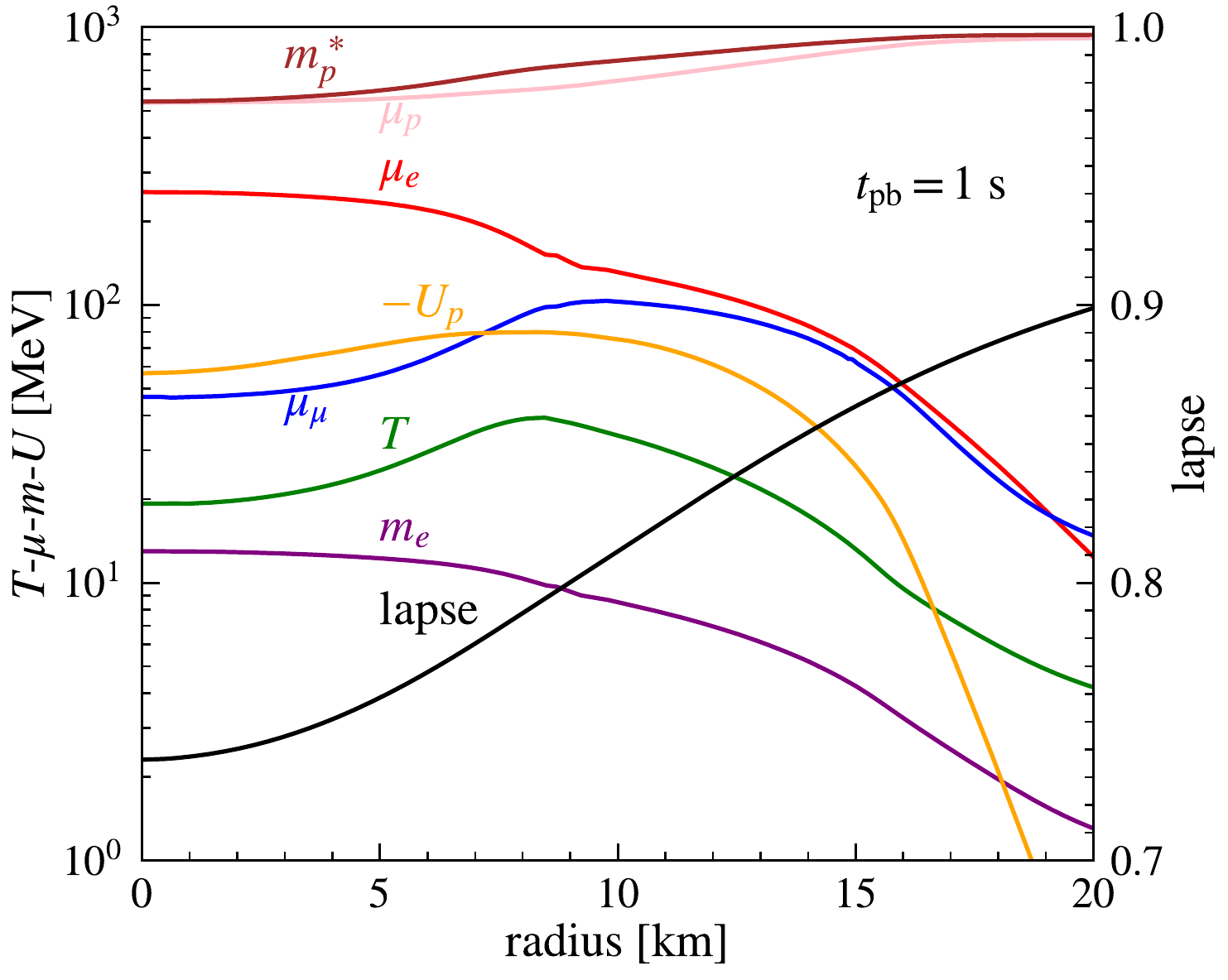}
\hspace{0.01\textwidth}
\includegraphics[width=0.48\textwidth]{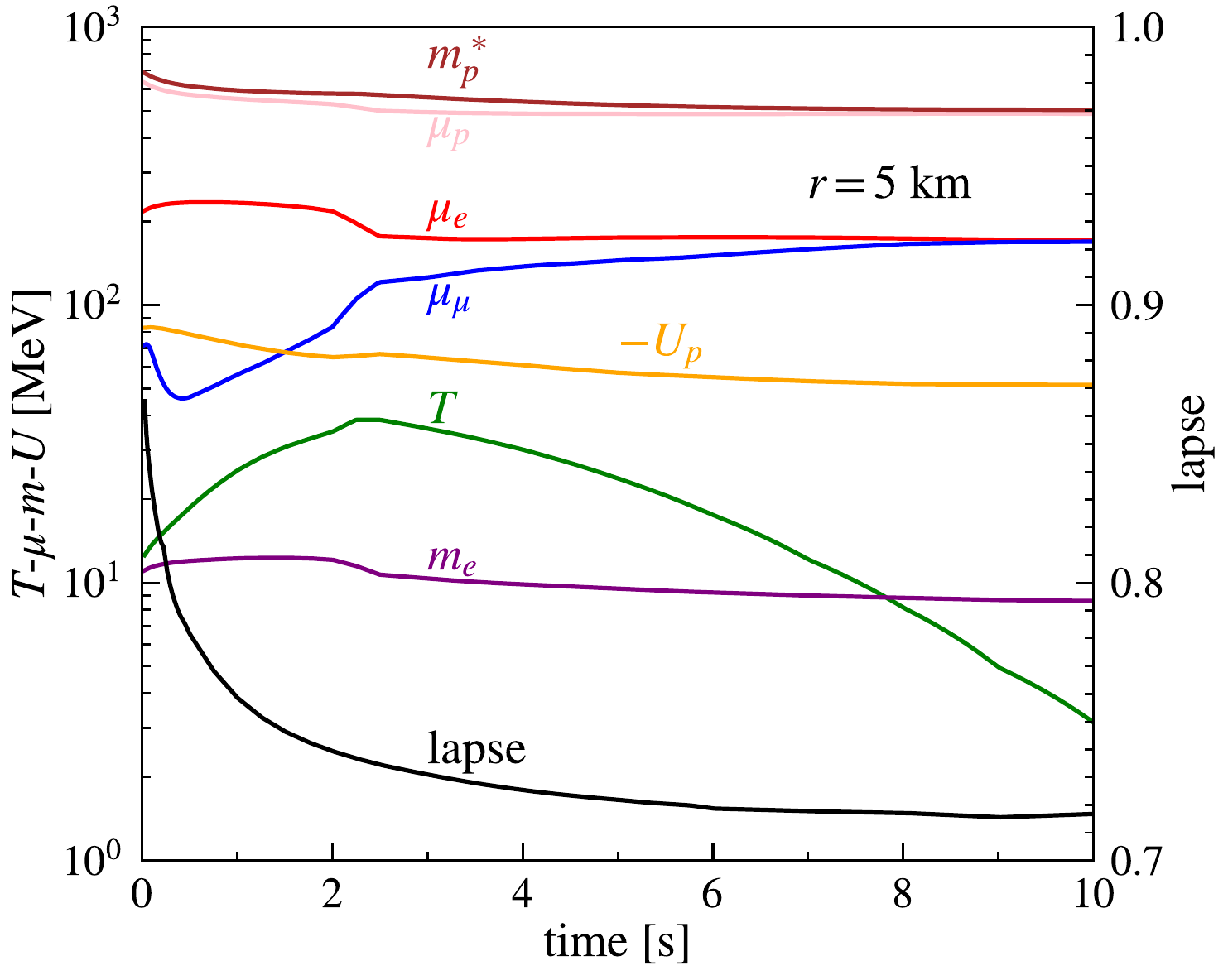}
\caption{\textbf{Left:}
Spatial evolution
of the Garching profiles for the SFHo-18.8 SN model \cite{garching-profile}
at 1 s post-bounce.
\textbf{Right:}
Temporal evolution of the 
same model
at a fixed radial distance of $r = 5$ km, spanning 10 seconds post-bounce. The displayed profiles include the temperature $T$, the electron chemical potential $\mu_e$, the muon chemical potential $\mu_\mu$, the negative of the proton interaction potential 
$U_p$, the in-medium proton mass $m_p^*$,
the proton chemical potential $\mu_p$ which includes $m_p^\ast$ \cite{Li:2025beu},
the gravitational lapse factor, and the in-medium electron mass $m_e$ calculated using Eq.~\eqref{eq:e-mass}.}
\label{fig:profiles}
\end{figure}

Due to 
plasma effects,
the electron mass in the SN core
is modified so that it deviates
significantly from its 
vacuum value,
$m_e^0=0.511$ MeV.
The in-medium electron mass is given by \cite{Braaten:1991hg,Lucente:2021hbp}:
\begin{equation}
    m_e = \frac{m_e^0}{\sqrt{2}} + \sqrt{\frac{(m_e^0)^2}{2} + \frac{\alpha}{\pi}(\mu_e^2 + \pi^2 T^2)},
    \label{eq:e-mass}
\end{equation}
where $\alpha = 1/137$ is the fine structure constant.

To compute the ALP production and absorption rates, 
we adopt the radial distributions and temporal evolution 
of various physical quantities 
from the SFHo-18.8 model \cite{Bollig:2020xdr,garching-profile}.
The left panel of
Fig.~\ref{fig:profiles} 
displays the spatial evolution of temperature $T$,
electron chemical potential $\mu_e$,
muon chemical potential $\mu_\mu$,
proton chemical potential $\mu_p$,  proton interaction potential $U_p$, in-medium proton mass $m_p^*$,
gravitational lapse factor,
and the in-medium electron mass $m_e$ calculated using Eq.~(\ref{eq:e-mass})
at $t_\mathrm{pb}=1$ s with $t_\mathrm{pb}$ denoting the post-bounce time.
The right panel 
of Fig.~\ref{fig:profiles}
shows the
time evolution of the same quantities 
at a fixed radial distance of $r = 5$ km,
spanning 10 seconds post-bounce.

\section{ALP production}
\label{sec:production}

\begin{figure}[htbp]\centering
\includegraphics[width=0.305\linewidth]{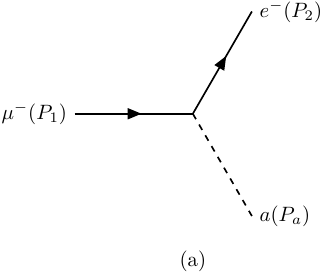}
\hspace{0.01\linewidth}
\includegraphics[width=0.32\linewidth]{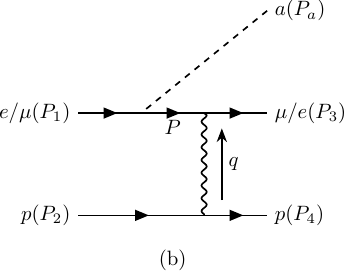}
\hspace{0.01\linewidth}
\includegraphics[width=0.32\linewidth]{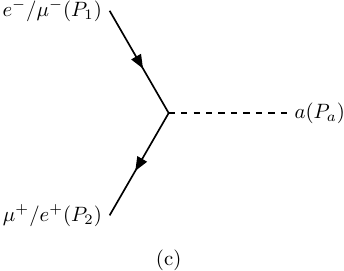}
\caption{Production processes of LFV-ALP in the SN:
(a) muon decay, 
(b) lepton bremsstrahlung, 
and 
(c) $e$-$\mu$ coalescence. 
} 
\label{fig:production:feynman}
\end{figure}

In this section we compute ALP production in the SN core 
and mantle
for the following four processes: 
(1) muon decay $\mu \to e+a$, 
(2) lepton bremsstrahlung $\mu(e) + p \to e(\mu) + p + a$, 
(3) electron-muon coalescence $e + \mu \to a$, 
and 
(4) semi-Compton $\mu(e) + \gamma \to e(\mu) + a$. 
While these processes are possible 
for both leptons and antileptons,  
we focus on lepton-induced reactions due to 
their significantly higher abundance in the SN core. 
The sole exception is electron-muon coalescence, 
which inherently requires both a lepton and an antilepton 
in the initial state. 
Fig.~\ref{fig:production:feynman}
shows the Feynman diagrams for 
muon decay, 
lepton bremsstrahlung, and 
electron-muon coalescence; 
Fig.~\ref{fig:FeynmanDiagram:Compton} shows  
the Feynman diagrams for semi-Compton. 
Note that muon decay is kinematically allowed only for ALP masses satisfying
$m_a<m_\mu-m_e$; 
electron-muon coalescence requires $m_a>m_\mu+m_e$.  
Lepton bremsstrahlung and semi-Compton have no  
such mass constraints. 
Also note that there are two important lepton bremsstrahlung processes: 
$e^- + p \to \mu^- + p + a$ and $\mu^- + p \to e^- + p + a$;  
following Ref.~\cite{Li:2025beu}, 
we consider the latter only for $m_\mu < m_a + m_e$. 
\footnote{In the mass range of $m_\mu > m_a + m_e$, 
the fermion (either electron or muon) propagator 
in the $\mu^- + p \to e^- + p + a$ process can be on-shell, 
if the virtual photon momentum goes to zero. 
In such cases, 
$\mu^- + p \to e^- + p + a$ effectively 
becomes the muon decay process.}

\subsection{Muon decay}

The Feynman diagram for the muon decay process
$\mu^-(P_1)\rightarrow e^-(P_2)+a(P_a)$ 
is shown in Fig.~\ref{fig:production:feynman} (a). 
The muon decay process can only occur if 
the mass condition 
$m_\mu>m_e+m_a$ 
is satisfied. 
The LFV-ALP production rate
per unit volume
per unit energy
is given by 
\cite{Li:2025beu}
\begin{equation}
\frac{d^2n_d}{dtdE_a}
=
\frac{|\mathcal{M}_d|^2}{32\pi^3}
\int_{E_2^-}^{E_2^+}
dE_2
f_\mu^-
(1-f_e^- ),
\end{equation}
where
$|\mathcal{M}_d|^2=2g_{ae\mu}^2[(m_\mu-m_e)^2-m_a^2]$
is the spin-summed squared matrix element,
$f_\ell^-=\left[e^{(E_\ell-\mu_\ell)/T}+1\right]^{-1}$ is 
the Fermi-Dirac distribution for fermion $\ell^-$,
and the integration limits are
\begin{equation}
E_2^\pm
=
\frac{E_a(m_\mu^2-m_e^2-m_a^2)}{2m_a^2}
\pm
\frac{\sqrt{E_a^2-m_a^2}}{2m_a^2}I(m_\mu,m_e,m_a),
\end{equation}
where
\begin{equation}
    I(x,y,z)=\sqrt{(x^2-y^2-z^2)^2-4y^2z^2}.
    \label{eq:Ifunction}
\end{equation}

\subsection{Lepton bremsstrahlung}

We now consider the lepton bremsstrahlung processes:
\begin{align}
e^-(P_1)+p(P_2) & \rightarrow\mu^-(P_3)+p(P_4)+a(P_a),\\
\mu^-(P_1)+p(P_2) & \rightarrow e^- (P_3)+p(P_4)+a(P_a),    
\end{align}
as shown in Fig.~\ref{fig:production:feynman} (b).
The ALP production rate
per unit volume 
per unit energy
is given by 
\cite{Lucente:2021hbp,Ferreira:2022xlw,Li:2025beu}
\begin{equation}
\frac{d^2n_b}{dtdE_a}
=
\frac{n_\mathrm{eff}|\textbf{p}_a|}{(2\pi)^632m_p^2}
\int_{m_3}^\infty
dE_3
\int_{-1}^1
dz_adz_3
\int_0^{2\pi}
d\phi
|\textbf{p}_1||\textbf{p}_3|
f_1(1-f_3)
|\mathcal{M}_b|^2,
\end{equation}
where 
$\phi$ is the angle between the
$\textbf{p}_1$-$\textbf{p}_a$ plane
and the
$\textbf{p}_1$-$\textbf{p}_3$ plane,
$z_a=\cos\theta_{1a}$
with
$\theta_{1a}$ being the angle
between
$\textbf{p}_1$ and $\textbf{p}_a$,
$z_3=\cos\theta_{13}$
with
$\theta_{13}$ being the angle
between
$\textbf{p}_1$ and $\textbf{p}_3$,
$n_\mathrm{eff}$ is the effective proton number density, and 
$\mathcal{M}_b$ is the matrix element.
Since the proton mass is much larger than 
the core temperature, $T\sim30$ MeV, 
proton recoil can be neglected. 
Following Refs.~\cite{Carenza:2021osu,Li:2025beu}, 
we adopt the static-proton approximation 
$E_2\simeq E_4\simeq m_p$, which yields  
$|\textbf{p}_1|=\sqrt{(E_3+E_a)^2-m_1^2}$.
The effective proton number density is given by 
\zb{\cite{Lucente:2021hbp,Ferreira:2022xlw}}
\begin{equation}
n_\mathrm{eff}=2
\int
\frac{d^3p_2}{(2\pi)^3}
f_p(E_2)
[1-f_p(E_2)], 
\label{eq:neff}
\end{equation}
where 
\begin{equation}
    f_p(E_p) = \frac{1}{e^{(E_p-\mu_p)/T}+1} 
    \label{eq:proton:occupation}
\end{equation} 
is the proton distribution function, 
with 
$E_p$ ($\mu_p$) being the proton energy (chemical potential). 
Taking into account plasma effects, 
the proton energy $E_p$ is given by  
\begin{equation}
E_p = \sqrt{m_p^{*2}+{\bf p}_2^2} + U_p, 
\end{equation}
where $m_p^{*}$ is the in-medium proton mass 
and $U_p$ is the proton interaction potential. 
Note that we have used 
the approximation $\mathbf{p}_2 \simeq \mathbf{p}_4$ \cite{Li:2025beu} 
in obtaining Eq.~\eqref{eq:neff}, 
and used Garching profiles \cite{garching-profile} 
for $m_p^{*}$, $\mu_p$, and $U_p$, 
which are shown in Fig.~\ref{fig:profiles}.
The matrix element for the lepton bremsstrahlung processes 
takes the form as
\begin{equation}
i\mathcal{M}_b
=
2e^2g_{ae\mu}m_p
\frac{1}{|\textbf{q}|\sqrt{\textbf{q}^2+k_s^2}}
\left[
\bar u_3\gamma^0
\frac{\slashed{P}+m_3}{P^2-m_3^2}\gamma_5u_1
+
\bar u_3\gamma_5
\frac{\slashed{Q}+m_1}{Q^2-m_1^2}\gamma^0u_1
\right],
\end{equation}
where
$P = P_1 - P_a$, 
$Q = P_3 + P_a$, and 
$k_s^2=(4\pi\alpha/T)\sum_jZ_j^2n_j$
is the Debye screening scale with
$n_j$ being the number density
of ion $j$ with charge $Z_je$
\cite{Raffelt:1985nk}.

\subsection{Electron-muon coalescence}

For ALPs with masses $m_a > m_\mu + m_e$, 
the electron-muon coalescence can occur, 
which consists of two processes: 
$e^-(P_1) + \mu^+(P_2) \to a (P_a)$ and 
$\mu^- (P_1) + e^+ (P_2) \to a (P_a)$, 
as shown in  Fig.~\ref{fig:production:feynman} (c). 
In both processes, we denote the four-momenta of the initial-state particles 
as $P_1$ (fermion) and $P_2$ (anti-fermion). 
The ALP production rate 
per unit volume 
per unit energy is
\cite{Li:2025beu}
\begin{equation}
\frac{d^2n_c}{dtdE_a}
=
\frac{|\mathcal{M}_c|^2}{32\pi^3}
\int_{E_2^-}^{E_2^+}
dE_2
\left(
f_\mu^- f_e^+ 
+f_e^- f_\mu^+
\right),
\end{equation}
with $|\mathcal{M}_c|^2 = 2g_{ae\mu}^2[m_a^2 - (m_\mu - m_e)^2]$, and 
$f_\ell^+=\left[e^{(E_\ell+\mu_\ell)/T}+1\right]^{-1}$ is 
the Fermi-Dirac distribution for anti-fermion $\ell^+$.
The integration limits $E_2^\pm$ for the anti-fermion energy are determined by:
\begin{equation}
    E_2^\pm =\frac{E_a(m_a^2-m_1^2+m_2^2)}{2m_a^2} \pm \frac{\sqrt{E_a^2-m_a^2}}{2m_a^2}I(m_1,m_2,m_a),
\end{equation}
where $m_1$ and $m_2$ represent the fermion and anti-fermion masses, respectively,
and $I$ is given by Eq.~\eqref{eq:Ifunction}.

\subsection{Semi-Compton scattering}

Recently, Ref.~\cite{Fiorillo:2025sln}
points out that semi-Compton scattering provide 
significant contributions to axions that couple to 
electrons. 
In this section,
we calculate the LFV-ALP production from 
the semi-Compton scattering.

\begin{figure}[htbp]
    \centering
    \includegraphics[width=0.65\linewidth]{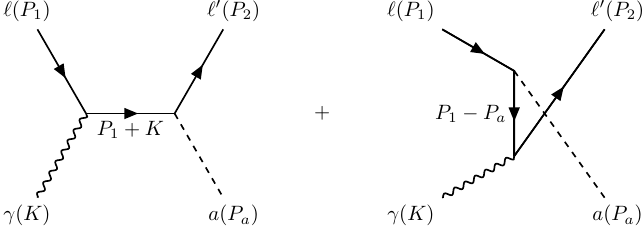}
    \caption{The Feynman diagrams for the semi-Compton scattering
    $\ell + \gamma\rightarrow \ell' + a$.}
    \label{fig:FeynmanDiagram:Compton}
\end{figure}

The matrix element
for the semi-Compton scattering
$\ell(P_1) + \gamma(K)\rightarrow\ell'(P_2) + a(P_a)$
shown in Fig.~\ref{fig:FeynmanDiagram:Compton}
is given by 
\begin{equation}
i\mathcal{M}_s
=
eg_{ae\mu}\epsilon_\lambda^\mu(K)\bar u(P_2)
\left(
\gamma_5
\frac{\slashed{P}_1+\slashed{K}+m_\ell}{s-m_\ell^2}\gamma_\mu
+
\gamma_\mu
\frac{\slashed{P}_1-\slashed{P}_a+m_{\ell'}}{u-m_{\ell'}^2}\gamma_5
\right)
u(p),
\end{equation}
where
$s\equiv(p_1+K)^2$ and
$u\equiv(P_1-P_a)^2$. 
The matrix element squared,
summed with the final state spins,
averaged with the initial state
spins and polarizations, is
\begin{equation}
\frac{1}{4}\sum_{\mathrm{spins},\lambda}
|\mathcal{M}_s|^2
=
e^2g_{ae\mu}^2
\left[
\frac{\langle\Phi_{ss}\rangle}{(s-m_\ell^2)^2}
+
\frac{\langle\Phi_{us}\rangle+\langle\Phi_{su}\rangle}{(s-m_\ell^2)(u-m_{\ell'}^2)}
+
\frac{\langle\Phi_{uu}\rangle}{(u-m_{\ell'}^2)^2}
\right],
\label{eq:SquaredMatrix}
\end{equation}
where
\begin{align}
\langle\Phi_{ss}\rangle
=&
m_\ell^2(2m_a^2+2m_\ell m_{\ell'}-3m_{\ell'}^2)
+s(m_{\ell'}^2-2m_\ell^2+2m_\ell m_{\ell'})+um_\ell^2-us,
\\
\langle\Phi_{su}\rangle
=&
\langle\Phi_{us}\rangle
=
(m_{\ell'}^2-m_\ell^2)(s-u)-us
-2m_\ell^2m_{\ell'}^2
\nonumber\\
&\qquad\qquad\qquad+
(m_a^2+m_\ell m_{\ell'})
\left[2(m_\ell^2+m_{\ell'}^2)-m_\ell m_{\ell'}-t\right],
\\
\langle\Phi_{uu}\rangle
=&
m_\ell^2(u-3m_{\ell'}^2)
+2m_\ell m_{\ell'}(m_{\ell'}^2+u)
+m_{\ell'}^2(2m_a^2+s-2u)-us. 
\end{align}
Here we have followed Ref.~\cite{Fiorillo:2025sln} by neglecting
the plasma mass correction 
to the initial photon and treated it as massless.
We note that Ref.~\cite{Li:2025beu} also carried out 
such an analysis, and our results agree with Ref.~\cite{Li:2025beu}; 
however, the cross term in Eq.~(\ref{eq:SquaredMatrix}) 
differs from Eq.~(20) of Ref.~\cite{Fiorillo:2025sln} 
by an overall minus sign. 
Such an error leads to 
an overestimation of the semi-Compton scattering 
in Ref.~\cite{Fiorillo:2025sln}.

The LFV-ALP production rate 
per unit volume 
per unit energy is
\cite{Fiorillo:2025sln}
\begin{align}
\frac{d^2n_s}{dtdE_a}
=
\int_0^\infty&\frac{\textbf{p}_1^2}{E_1}d|\textbf{p}_1|
\int_{\omega_p}^\infty|\textbf{k}|d|\textbf{k}|
\int_{-1}^1d\mu
\int_0^{2\pi}d\phi
\Theta(1-|\cos\xi|)
\nonumber\\
&\times
\frac{f_\ell(E_1)f_\gamma(|\textbf{k}|)\left[1-f_{\ell'}(E_1+|\textbf{k}|-E_a)\right]}{512\pi^6|\textbf{p}_1+\textbf{k}|}\sum_{\mathrm{spins},\lambda}|\mathcal{M}_s|^2,
\label{eq:Boltzmann:Compton}
\end{align}
where 
$f_\gamma \left(|\textbf{k}| \right) = \left[ \exp \left(|\textbf{k}|/T \right) - 1 \right]^{-1}$ is the Bose-Einstein distribution for photons,
$\mu=\cos\theta$ with $\theta$
being the angle between $\textbf{p}_1$ and $\textbf{k}$,
and 
$\xi$ ($\phi$) is the ALP polar (azimuthal) angle
with respect to $\mathbf{p}_1+\mathbf{k}$. 
The delta function
associated with energy conservation leads to
\begin{equation}
\cos\xi=
\frac{
	\textbf{p}_1\cdot\textbf{k}
	+
	E_a E_1
	+
	(E_a-E_1)
	|\textbf{k}|
	+(m_{\ell'}^2-m_\ell^2-m_a^2)/2
}{|\textbf{p}_1+\textbf{k}||\textbf{p}_a|}. 
\label{ConstraintsonX}
\end{equation}
For the lower integration limit of $|\textbf{k}|$ 
in Eq.~(\ref{eq:Boltzmann:Compton}), 
we use the plasma frequency $\omega_p$ in the relativistic limit, 
which is given by
\cite{Braaten:1993jw}
\begin{equation}
\omega_p^2
=
\frac{4\alpha}{3\pi}
\left(
\mu_e^2
+
\frac{1}{3}\pi^2T^2
\right).
\end{equation}
We note that $\omega_p$ effectively removes the divergence 
in the $|\textbf{k}|\rightarrow0$ regime.

By solving 
$|\cos\xi|\leq1$,
we derive the constraints on $\mu$
from Eq.~(\ref{ConstraintsonX}) as
\begin{equation}
	\mu_{\min}\equiv\frac{X_1-X_2}{|\textbf{p}_1||\textbf{k}|}
\leq\mu\leq
\frac{X_1+X_2}{|\textbf{p}_1||\textbf{k}|}
\equiv\mu_{\max},
\end{equation}
where
\begin{align}
X_1 & \equiv(E_a-E_1)(E_a-|\textbf{k}|)+(m_\ell^2-m_{\ell'}^2-m_a^2)/2,
\\
X_2 & \equiv|\textbf{p}_a|\sqrt{(E_1+|\textbf{k}|-E_a)^2-m_{\ell'}^2}.
\end{align}
Note that
the square root in $X_2$
is always valid since 
$E_1\geq E_a+m_{\ell'}-|\textbf{k}|$
is guaranteed by the energy conservation.
Thus, 
the LFV-ALP production rate
per unit volume
per unit energy
is given by
\begin{align}
	\frac{d^2n_s}{dtdE_a}
=
\int_0^\infty&\frac{\textbf{p}_1^2}{E_1}
d|\textbf{p}_1|
\int_{k_{\min}}^\infty
|\textbf{k}|d|\textbf{k}|
\int_{\mu_{\min}}^{\mu_{\max}}
d\mu\Theta(1-|\mu|)
\int_0^{2\pi}\mathrm{d}\phi
\nonumber\\
&\times
\frac{f_\ell(E_1)f_\gamma(|\textbf{k}|)\left[1-f_{\ell'}(E_1+|\textbf{k}|-E_a)\right]}{512\pi^6|\textbf{p}_1+\textbf{k}|}\sum_{\mathrm{spins},\lambda}|\mathcal{M}_s|^2,
\label{eq:Boltzmann:Compton:2}
\end{align}
where
$k_{\min}=\max\{
E_a+m_{\ell'}-E_1,
\omega_p
\}$.
Although
Eqs.~(\ref{eq:Boltzmann:Compton})
and (\ref{eq:Boltzmann:Compton:2})
are identical analytically, 
we prefer to use
Eq.~(\ref{eq:Boltzmann:Compton:2})
for numerical calculations
since the Heaviside function here
only constrains a single
integration variable
$\mu$ explicitly,
unlike the Heaviside function
in Eq.~(\ref{eq:Boltzmann:Compton}),
which constrains the
three integration variables
$|\textbf{p}_1|$,
$|\textbf{k}|$,
and $\mu$
implicitly.

To compute the production rate, we express 
the Mandelstam variables 
$s$ and $t$
in terms of the integration variables: 
\begin{align}
&s
=(P_1+K)^2
=
m_\ell^2
+2|\textbf{k}|(E_1-|\textbf{p}_1|\mu),
\\
&t
=(K-P_a)^2=m_a^2+2|\textbf{k}|
\left[
|\textbf{p}_a|
(\cos\xi\cos\zeta-\sin\xi\cos\phi\sin\zeta)
-E_a
\right],
\end{align}
where   
$\cos\zeta = 
(|\textbf{p}_1|\mu+|\textbf{k}|)
/{|\textbf{p}_1+\textbf{k}|}$, and 
$\sin\zeta =
|\textbf{p}_1|\sqrt{1-\mu^2}/{|\textbf{p}_1+\textbf{k}|}$.
The derivation of these expressions is provided in 
appendix \ref{sec:ALP:p:semi-compton}. 
We then determine the Mandelstam variable $u$ via
\begin{equation}
u=m_\ell^2+m_{\ell'}^2+m_a^2-s-t.
\end{equation}

\subsection{Comparison between different production channels}

\begin{figure}[htbp]
\centering
\includegraphics[width=0.5\textwidth]{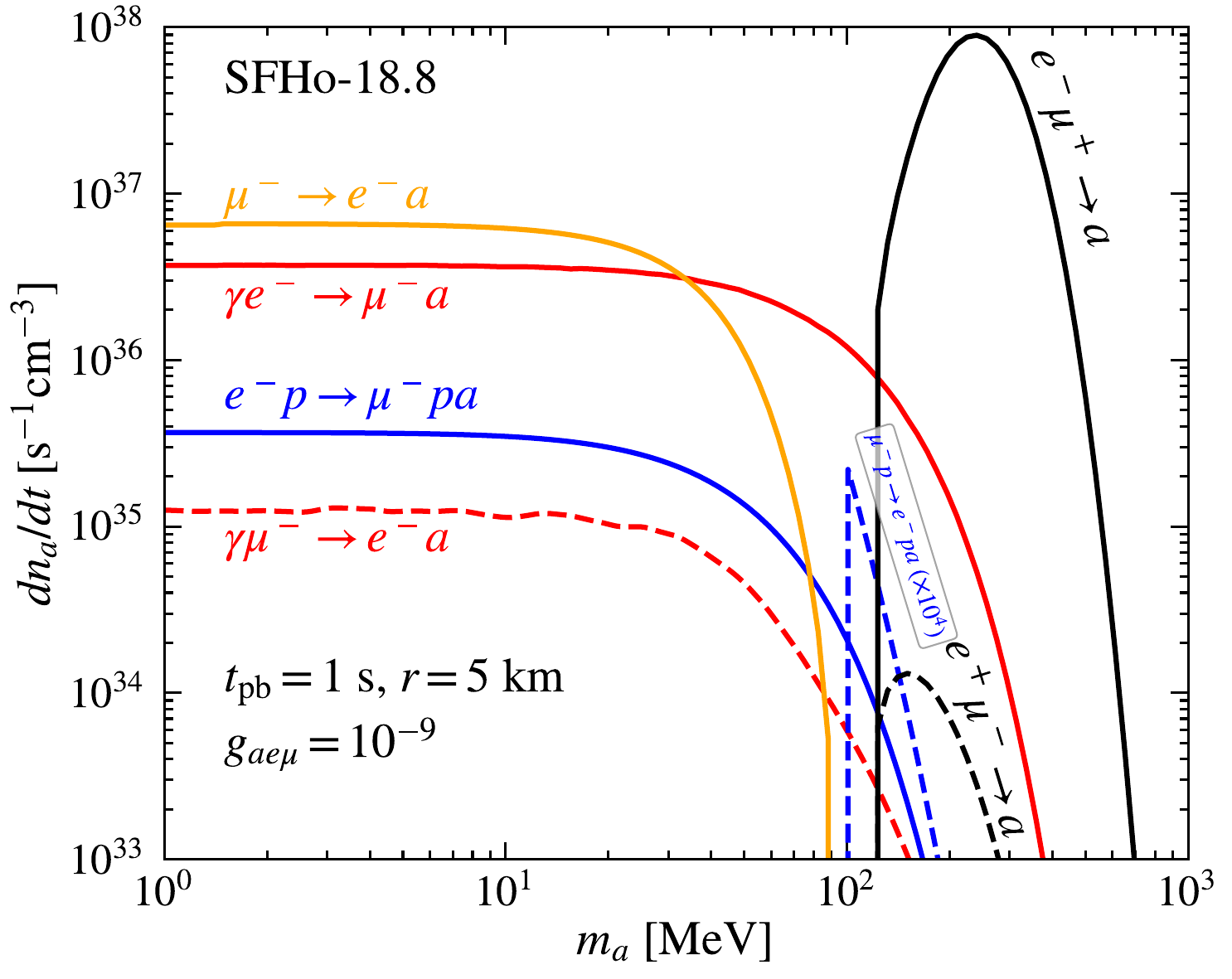}
\caption{LFV-ALP production rates per unit volume 
at $t_\mathrm{pb}=1$ s and $r=5$ km
with $g_{ae\mu}=10^{-9}$ 
from various channels: 
(1) muon decay $\mu^-\rightarrow e^-a$ (solid orange),
(2) lepton bremsstrahlung $e^-p\rightarrow\mu^-pa$ (solid blue)
and $\mu^-p\rightarrow e^-pa$ (dashed blue),  
(3) electron-muon coalescence 
$e^-\mu^+\rightarrow a$ (solid black)
and $e^+\mu^-\rightarrow a$ (dashed black), 
and 
(4) semi-Compton scattering $\gamma e^-\rightarrow\mu^-a$ (solid red)
and $\gamma\mu^-\rightarrow e^-a$ 
(dashed red). 
We use the Garching group's muonic model SFHo-18.8. 
The $\mu^-p\rightarrow e^-pa$ channel 
is shown for masses above $\sim 100$ MeV 
(multiplied with a factor of $10^4$)  
since we require $m_e+m_a>m_\mu$ for this channel 
to avoid double counting with muon decay.} 
\label{fig:comparison:productionratesperunitVol}
\end{figure}

To compare different production channels, 
we compute the LFV-ALP production rate per unit volume via 
\begin{equation}
\frac{dn_a}{dt} = \int_{m_a}^{\infty}
dE_a \frac{dn_a}{dt dE_a}, 
\label{eq:productionrateperV}
\end{equation}
where ${dn_a}/{dt dE_a}$ is the 
production rate per unit volume per unit energy. 
Fig.~\ref{fig:comparison:productionratesperunitVol} 
shows the LFV-ALP production rates per unit volume  
from different production channels 
at one second post-bounce and $r=5$ km
with $g_{ae\mu}=10^{-9}$.
The muon decay process $\mu^-\rightarrow e^-a$ dominates
for $m_a\lesssim30$ MeV, 
the electron-muon coalescence process  $e^-\mu^+\rightarrow a$ dominates
for $m_a\gtrsim110$ MeV,
and the semi-Compton scattering process $e^-\gamma\rightarrow\mu^-a$
dominates in the intermediate mass range 
$30\ \mathrm{MeV}\lesssim m_a\lesssim 110$ MeV. 
The lepton bremsstrahlung process is always subdominant. 
For the lepton bremsstrahlung,
electron-muon coalescence, and 
semi-Compton scattering processes, 
the production rates involving an initial electron are greater
than those involving an initial muon,
due to the higher electron number density.

\begin{figure}
    \centering
    \includegraphics[width=0.46\linewidth]{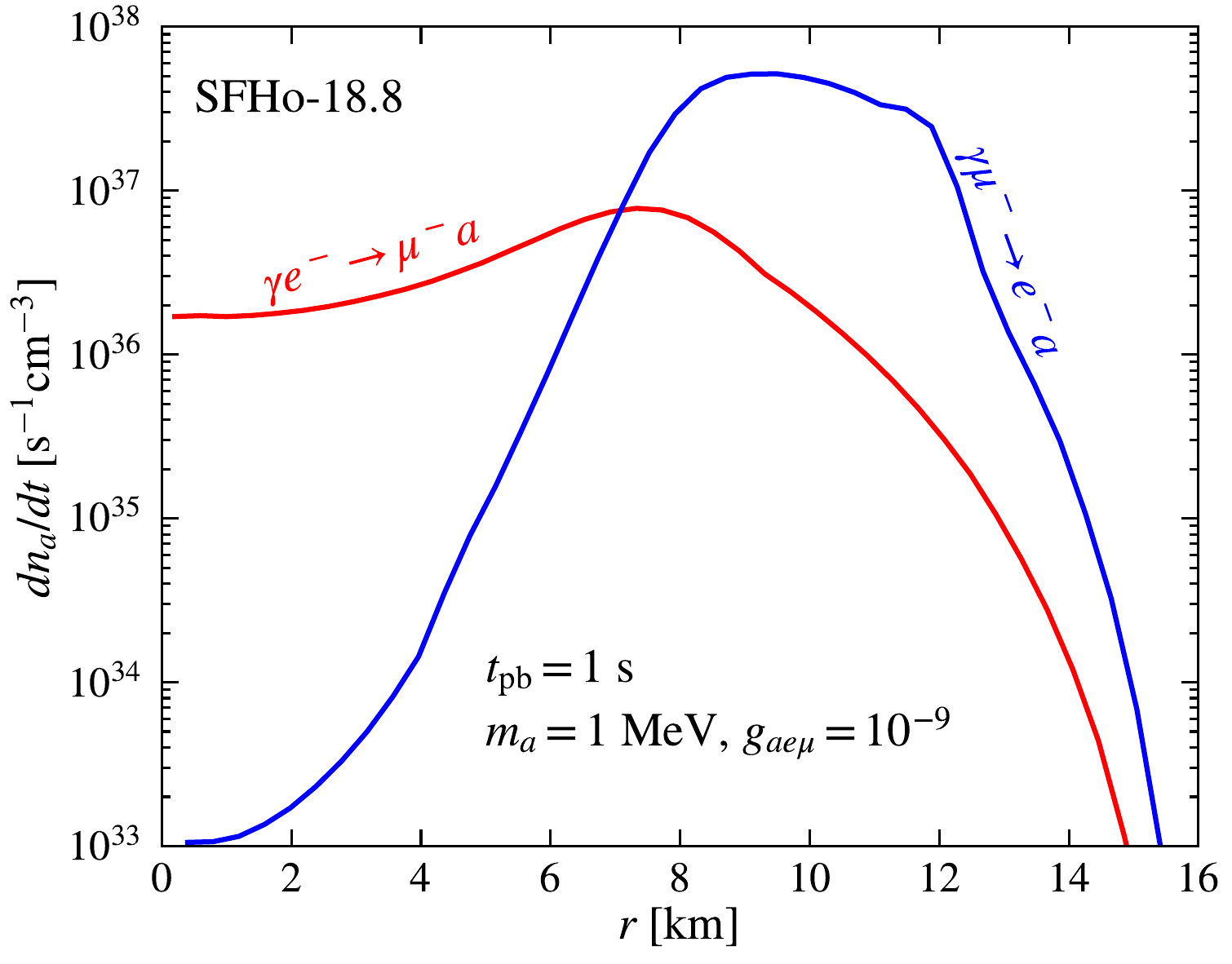}
    \includegraphics[width=0.45\linewidth]{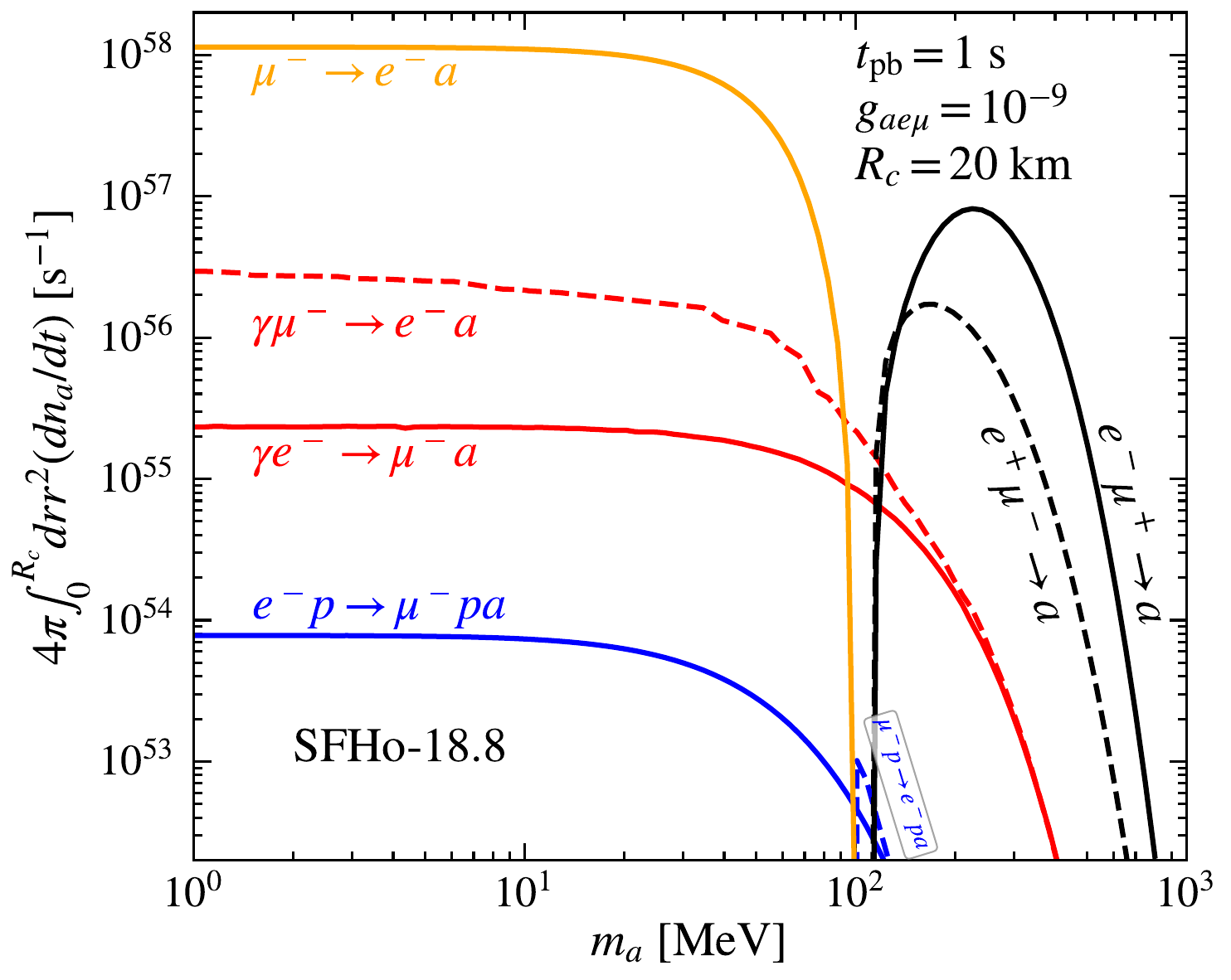}
    \caption{
\textbf{Left panel:}
The LFV-ALP production rate per unit volume
from the
semi-Compton scattering processes
$\gamma e^-\rightarrow \mu^-a$ (red)
and
$\gamma\mu^-\rightarrow e^-a$ (blue)
varying with the radial coordinate
at $t_\mathrm{pb}=1$ s
for $m_a=1$ MeV 
and
$g_{ae\mu}=10^{-9}$.
\textbf{Right panel:}
The total LFV-ALP production rates 
within the SN core, $4\pi \int_0^{R_c} dr r^2 (dn_a/dt)$,
at $t_\mathrm{pb}=1$ s
with $g_{ae\mu}=10^{-9}$
from various channels:
(1) muon decay $\mu^-\rightarrow e^-a$ (solid orange),
(2) lepton bremsstrahlung $e^-p\rightarrow\mu^-pa$ (solid blue)
and $\mu^-p\rightarrow e^-pa$ (dashed blue),  
(3) electron-muon coalescence 
$e^-\mu^+\rightarrow a$ (solid black)
and $e^+\mu^-\rightarrow a$ (dashed black), 
and 
(4) semi-Compton scattering $\gamma e^-\rightarrow\mu^-a$ (solid red)
and $\gamma\mu^-\rightarrow e^-a$ 
(dashed red). 
We use the 
Garching group's muonic model SFHo-18.8.
}
\label{fig:comparison:productionrate}
\end{figure}

We also plot the production rate per unit volume
from the semi-Compton scattering processes
varying with the radial coordinate $r$
at $t_\mathrm{pb}=1$ s,
as shown in the left panel of Fig.~\ref{fig:comparison:productionrate},
where we fix
$m_a=1$ MeV and $g_{ae\mu}=10^{-9}$.
We find that the LFV-ALP production rate per unit volume from
$\gamma e^- \rightarrow \mu^- a$
dominates over that from
$\gamma \mu^- \rightarrow e^- a$
for $r \lesssim 7~\mathrm{km}$,
where the electron and muon chemical potentials are comparable.
Conversely, for $r \gtrsim 7~\mathrm{km}$,
the production rate from
$\gamma \mu^- \rightarrow e^- a$
becomes dominant.

We further compute the total LFV-ALP production rates 
within the SN core via 
\begin{equation}
4\pi \int_0^{R_c} dr r^2 \frac{dn_a}{dt}, 
\end{equation}
where $dn_a/dt$ is given in Eq.~(\ref{eq:productionrateperV}). 
The right panel of Fig.~\ref{fig:comparison:productionrate}
shows the total LFV-ALP production rates 
within the SN core
at $t_\mathrm{pb}=1$ s 
with $g_{ae\mu}=10^{-9}$ 
from various channels. 
The muon decay process $\mu^- \rightarrow e^- a$ dominates for $m_a \lesssim 90$ MeV, 
the electron–muon coalescence process $e^- \mu^+ \rightarrow a$ dominates for $m_a \gtrsim 110$ MeV, 
and the semi-Compton scattering process $\gamma \mu^- \rightarrow e^- a$ 
dominates in the intermediate mass range $90~\mathrm{MeV} \lesssim m_a \lesssim 110~\mathrm{MeV}$. 
The lepton bremsstrahlung process remains subdominant throughout the entire mass range. 
For electron–muon coalescence, 
the production rate involving an initial electron exceeds that with an initial muon, 
while for lepton bremsstrahlung and semi-Compton scattering, the rates involving an initial electron 
are likewise higher than those involving an initial muon.

\section{ALP absorption rate}
\label{sec:absorption}

\begin{figure}[htbp]\centering
\includegraphics[width=0.3\linewidth]{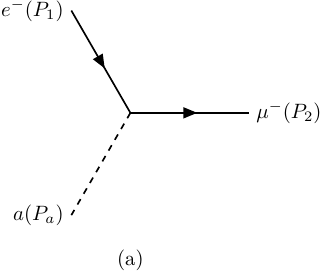}
\hspace{0.01\linewidth}
\includegraphics[width=0.32\linewidth]{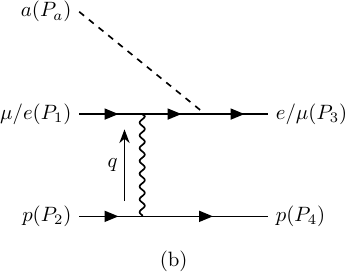}
\hspace{0.01\linewidth}
\includegraphics[width=0.32\linewidth]{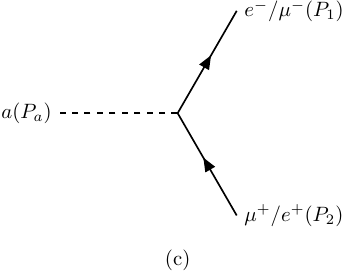}
\caption{Absorption processes of LFV-ALP in the SN: 
(a) electron-ALP coalescence, 
(b) inverse bremsstrahlung, 
and 
(c) ALP decay, 
which are the inverse processes of 
the three production processes 
shown in Fig.~\ref{fig:production:feynman}.}
\label{fig:absorption:feynman}
\end{figure}

\begin{figure}[htbp]
    \centering
    \includegraphics[width=0.65\linewidth]{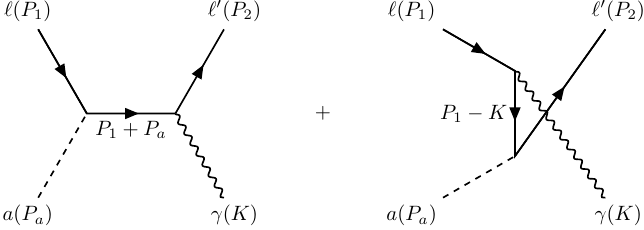}
    \caption{The Feynman diagrams for the inverse semi-Compton scattering
    $a + \ell \rightarrow\ell' + \gamma$, 
    which are the inverse processes of the semi-Compton scattering shown in 
    Fig.~\ref{fig:FeynmanDiagram:Compton}.}
    
    \label{fig:FeynmanDiagram:InverseCompton}
\end{figure}

In this section we compute the ALP absorption rate. 
Figs.~\ref{fig:absorption:feynman}
and \ref{fig:FeynmanDiagram:InverseCompton} 
show the ALP absorption processes: 
(1) electron-ALP coalescence, 
(2) inverse bremsstrahlung,  
(3) ALP decay, and 
(4) inverse semi-Compton. 
These ALP absorption processes  
are the inverse processes of 
the ALP production processes, 
as shown in 
Figs.~\ref{fig:production:feynman}
and \ref{fig:FeynmanDiagram:Compton}.

The ALP absorption rate can be 
obtained from the production rate of the inverse process  
as
\cite{Li:2025beu}
\begin{equation}
\Gamma_A(E_a)
=
e^{(E_a-\mu_a^0)/T}
\Gamma_E
=
e^{(E_a-\mu_a^0)/T}
\frac{2\pi^2}{|\textbf{p}_a|E_a}
\frac{d^2n_a}{dtdE_a},
\end{equation}
where 
$E_a=\sum_iE_i-\sum_{j\neq a}E_j$
and 
$\mu_a^0\equiv
\sum_i\mu_i
-
\sum_{j\neq a}\mu_j$
with $i$ ($j$) denoting the final (initial) particles in the absorption process.
We note that the reduced absorption rate 
can be determined solely by the absorption rate via 
$\Gamma = \Gamma_A - \Gamma_E = \left[1 - e^{-(E_a - \mu_a^0)/T}\right] \Gamma_A$.

In the large mass region $m_a > m_\mu + m_e$, 
the dominant absorption process is 
$a\rightarrow e^\mp\mu^\pm$ (ALP decay). 
In this case, 
for the absorption term up to the progenitor radius
(the second term in brackets in Eq.~\eqref{eq:energy-deposition}), 
we approximate the ALP decay rate with 
its vacuum value: 
\begin{equation}
\Gamma_A^\mathrm{vac} = \frac{g_{ae\mu}^2 
[m_a^2 - (m_\mu - m_e^0)^2]}{4\pi m_a^2 E_a}I(m_a,m_\mu,m_e^0). 
\end{equation} 
This is because 
both temperature and mass density 
decline steeply beyond the core such that  
ALPs effectively decay in vacuum.

\section{LESN constraints}
\label{sec:results}

We compute the LESN constraints on LFV-ALPs by demanding that 
the net energy deposition $E_d$ in the SN mantle, 
calculated via Eq.~\eqref{eq:energy-deposition-total}, 
does not exceed the LESN explosion energy: 
$E_d \le 0.1$ B. 
We consider four ALP production processes: 
(1) muon decay, 
(2) lepton bremsstrahlung,
(3) electron-muon coalescence, and
(4) semi-Compton scattering.

\begin{figure}[htbp]
\centering
\includegraphics[width=0.5\textwidth]{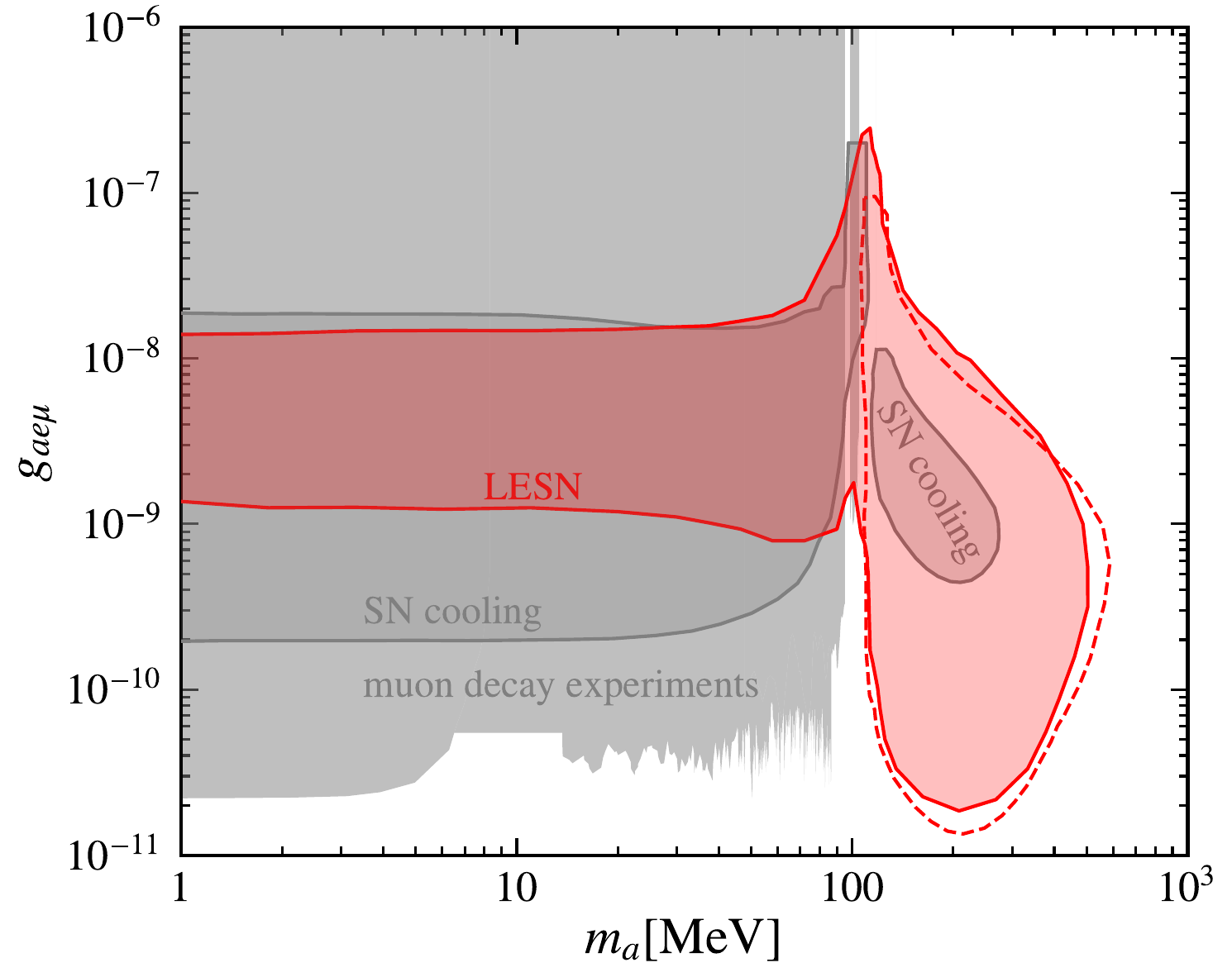}
\caption{The LESN constraints on LFV-ALPs for $E_d \le 0.1$ B
(red-shaded). 
Also shown are the LESN constraints  
derived in Ref.~\cite{Huang:2025rmy} (red-dashed). 
Gray-shaded regions are constraints 
derived from SN cooling bounds \cite{Li:2025beu} 
and rare muon decay experiments 
\cite{Derenzo:1969za,Bilger:1998rp,Jodidio:1986mz,TWIST:2014ymv,PIENU:2020loi}, 
which are taken from Ref.~\cite{Li:2025beu}.}
\label{fig:ALPlimits-with-mantle}
\end{figure}

Fig.~\ref{fig:ALPlimits-with-mantle} 
shows the LESN constraints on LFV-ALPs, 
derived under the condition $E_d \le 0.1$~B. 
We also show existing bounds, 
including those from rare muon decay experiments 
\cite{Derenzo:1969za,Bilger:1998rp,Jodidio:1986mz,TWIST:2014ymv,PIENU:2020loi} 
and SN cooling \cite{Li:2025beu}.

In the low-mass region ($m_a \lesssim 100$~MeV), 
the most stringent constraints arise from rare muon decay experiments 
\cite{Derenzo:1969za,Bilger:1998rp,Jodidio:1986mz,TWIST:2014ymv,PIENU:2020loi}, 
which probe the coupling $g_{ae\mu}$ down to the level of $\sim 10^{-11}$. 
The next most stringent constraints come from SN cooling limits \cite{Li:2025beu}, 
while the LESN bounds reach only $g_{ae\mu} \sim 10^{-9}$, 
about one order of magnitude weaker. 
This outcome runs contrary to the naive expectation based on the energy budget, 
since the LESN explosion energy ($\sim 0.1$~B) is approximately  
three orders of magnitude smaller than the upper bound on the 
ALP energy escaping from the SN core in the cooling analyses 
($\sim 100$~B), 
suggesting that the LESN energy budget should, 
in principle, lead to a tighter constraint. 
The weaker LESN limits instead reflect 
the inefficient absorption of ALPs in the mantle. 
In the low-mass region, the relevant absorption processes
(namely electron–ALP coalescence, 
inverse bremsstrahlung,
and inverse semi-Compton scattering) 
depend on the electron or muon number densities,
both of which decrease rapidly with radius outside the SN core.
Consequently, most ALPs produced in the core escape the progenitor star
without being reabsorbed, leading to weaker LESN limits despite the 
tighter energy budget.

For ALPs with masses above the kinematic threshold of 
$m_a > m_\mu + m_e$, 
the muon decay channel is kinematically forbidden, 
and the corresponding rare muon decay limits are 
therefore inapplicable. 
In this mass range, SN observations 
provide the most stringent constraints, 
and 
the dominant ALP production channel in the SN 
is the electron-muon coalescence process. 
We find that LESNe probe a broader region of parameter space 
than the SN cooling limits, 
which is opposite to the trend observed in the low-mass regime. 
This difference arises mainly because, in the high-mass region,  
the dominant ALP absorption channel is ALP decay, 
whose rate increases with the radial coordinate $r$ 
in the SN mantle, 
leading to highly efficient absorption 
in this region. 
Our current analysis shows that LESNe can probe 
the $g_{ae\mu}$ coupling  
down to $\sim 2 \times 10^{-11}$ at $m_a \simeq 200$ MeV, 
which is about a factor of two weaker than 
the previous LESN result in Ref.~\cite{Huang:2025rmy}, 
shown as the red-dashed contour 
in Fig.~\ref{fig:ALPlimits-with-mantle}. 
This difference arises mainly because the present analysis employs 
the Garching SN profiles spanning the full 
ten-seconds post-bounce evolution, whereas 
Ref.~\cite{Huang:2025rmy} used only 
the one-second post-bounce profile. 
In addition, the current analysis adopts 
a more accurate treatment of ALP absorption, 
by performing explicit trajectory integrations.

For ALPs with masses in the range 
$m_\mu-m_e< m_a < m_\mu+m_e$, 
both muon decay and 
electron-muon coalescence are kinematically forbidden. 
In this intermediate mass region, 
the dominant production channel is the semi-Compton scattering process, 
as shown in Fig.~\ref{fig:comparison:productionratesperunitVol}. 
Since the in-medium electron mass is $\sim$13~MeV at $r=20$~km 
and decreases to $\sim$1~MeV at $r=50$~km, 
this mass window corresponds approximately to $m_a\simeq (100, 110)$~MeV. 
Within this range, we find that 
the LESNe provide the most stringent constraints, 
excluding the $g_{ae\mu}$ coupling down to $\sim  10^{-9}$.

\subsection{Energy drain in the mantle}

\begin{figure}[htbp]
\centering
\includegraphics[width=0.5\textwidth]{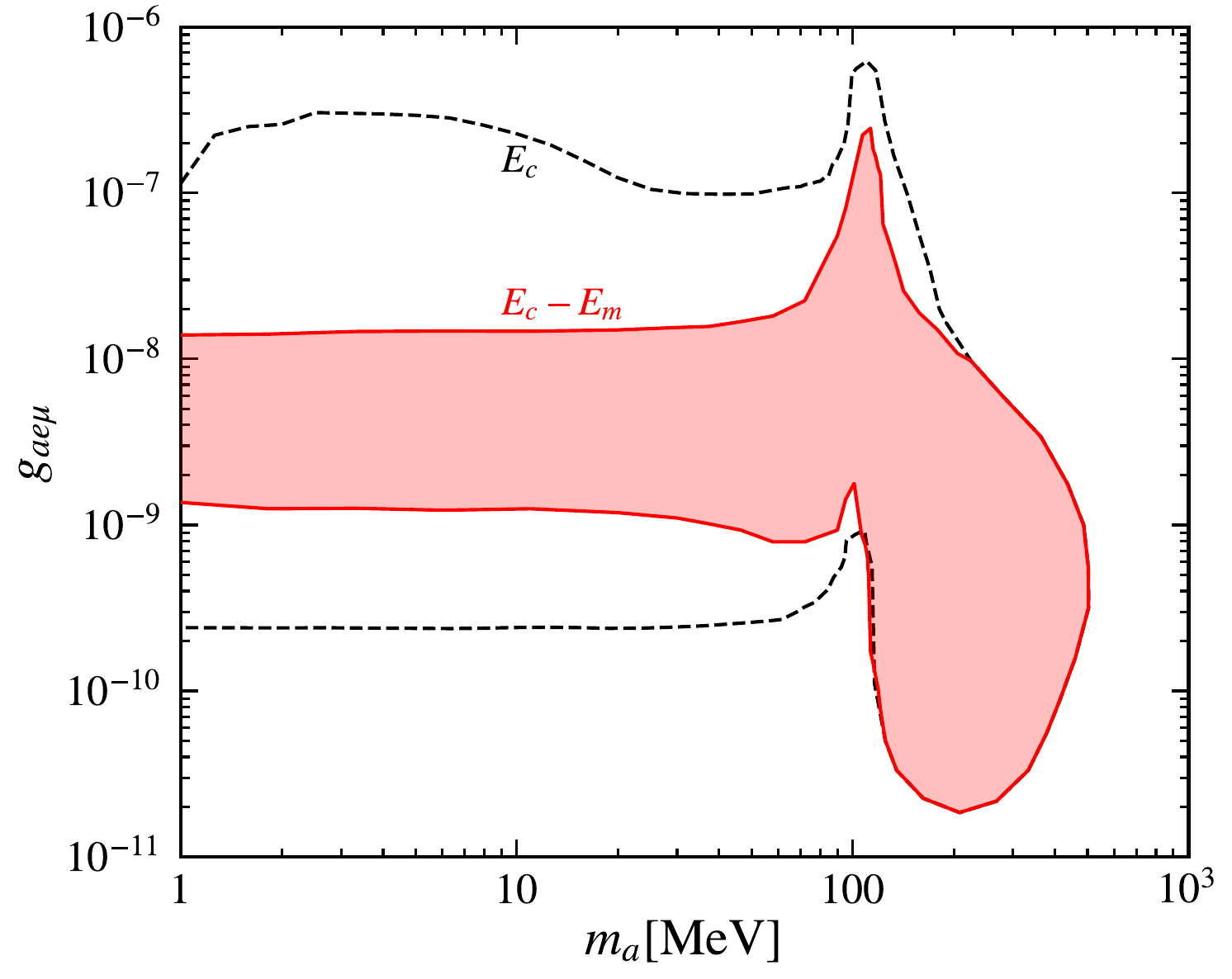}
\caption{Comparison of LESN constraints 
with (red-solid) and without (black-dashed) 
accounting for the energy drain in the mantle.}
\label{fig:energy:drain:comparison}
\end{figure}

In Fig.~\ref{fig:energy:drain:comparison},  
we compare the LESN constraints obtained 
with and without accounting for the energy drain in the mantle, 
the $E_m$ term in Eq.~\eqref{eq:energy-deposition-total}. 
We find that 
neglecting this contribution leads to a much larger excluded parameter space. 
We find that for strong couplings, the inclusion of the $E_m$ term 
has a pronounced impact and can modify the constraints 
on $g_{ae\mu}$ by more than an order of magnitude. 
Interestingly, for ALP masses below $\sim 100$~MeV, 
the effects of the $E_m$ term remain substantial 
even in the weak-coupling regime, a feature that, 
to our knowledge, has not been emphasized in previous studies.  
This behavior arises because, for $m_a\lesssim m_\mu+m_e$, 
the ALP decay process is kinematically forbidden, allowing  
the LFV-ALPs produced in the mantle to escape 
the progenitor star nearly freely, thereby causing 
a significant energy loss. 
Overall, these findings highlight the importance of properly 
including the mantle energy-drain effect  
in deriving reliable LESN constraints on LFV-ALPs 
for both strong and weak couplings.

We note that in the regime of large couplings, 
LFV-ALPs can reach equilibrium with the medium. 
In this case, the requirement of a vanishing ALP chemical potential 
implies $\mu_e=\mu_\mu$. 
In our present analysis, however, we use 
the Garching group's muonic model SFHo-18.8, 
where the electron and muon chemical potentials are not equal. 
A more accurate determination of the limits in the strong-coupling 
regime would therefore require a more sophisticated 
treatment that self-consistently accounts for these effects,
which we leave for future work.

\section{Summary}
\label{sec:summary}

In this paper, we present refined
LESN constraints on LFV-ALPs.
We consider four production channels
in the SN:
muon decay,
lepton bremsstrahlung,
electron-muon coalescence,
and semi-Compton scattering, 
the last of which is analyzed here for the first time for LFV-ALPs in the SN. 
We find that muon decay and electron-muon coalescence dominate 
the ALP production in the small and large mass regimes, respectively, 
and semi-Compton scattering provides 
leading contributions in the intermediate mass range. 
The lepton bremsstrahlung process is found to be subdominant.
We compute the net energy deposition
in the SN mantle 
by taking into account both the 
energy transfer from the core to the mantle 
and the energy drain from the mantle due to 
ALP production within it. 
We find that the energy drain from the mantle are substantical 
for both strong- and weak-coupling regimes, the latter of which, 
to our knowledge, has not been noted in the literature. 
By requiring the net energy deposited in the
SN mantle to be less than the LESN explosion energy (0.1 B), 
and by properly accounting for ALP absorption
using the Garching group's muonic model SFHo-18.8, 
which spans about ten seconds post-bounce, 
we obtain more accurate LESN constraints. 
We find that LESNe can provide the most stringent constraints 
on LFV-ALPs with masses above $\sim$110 MeV.

\begin{acknowledgments}
We thank 
Yonglin Li, Wenxi Lu, and Zicheng Ye 
for discussions. 
We thank Hans-Thomas Janka 
for providing the SN profiles 
used for numerical calculations. 
The work is supported in part by the 
National Natural Science Foundation of China under Grant 
No.\ 12275128. 
\end{acknowledgments}

\appendix
\section{ALP momentum}
\label{sec:ALP:p:semi-compton}

\begin{figure}[htbp]
    \centering
    \includegraphics[width=0.5\linewidth]{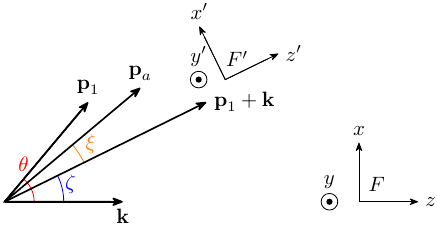}
    \caption{The momentum directions of the initial charged lepton ($\textbf{p}_1$) and photon ($\textbf{k}$),
		and the final ALP ($\textbf{p}_a$).}
    \label{fig:MomentumDirection}
\end{figure}

In this section we express
the Mandelstam variables in the semi-Compton scattering
in terms of the integration variables. We work in the frame $F$ where 
the $z$-axis is along the   
momentum of the initial photon ${\bf k}$ 
and the momentum of the initial charged lepton 
${\bf p}_1$ is
in the $xz$-plane. In the frame $F$ we have
\begin{equation}
K^\mu=|\textbf{k}|(1,0,0,1),
\quad
P_1^\mu=(E_1,|\textbf{p}_1|\sin\theta,0,|\textbf{p}_1|\cos\theta). 
\end{equation}
where $\theta$ is the polar angle of ${\bf p}_1$.
To obtain the ALP momentum, 
we work in another frame $F'$
where the $z'$-axis is along 
$\textbf{p}_1+\textbf{k}$ 
and the $y'$-axis is the same as the $y$-axis 
in the $F$ frame, 
as shown in Fig.~\ref{fig:MomentumDirection}. 
In the frame $F'$, we have
\begin{equation}
	P_a'^\mu=
	(E_a,|\textbf{p}_a|\sin\xi \cos\phi,|\textbf{p}_a|\sin\xi \sin\phi,|\textbf{p}_a|\cos\xi),
\end{equation} 
where $\xi(\phi)$ is the ALP polar (azimuthal) angle in the $F'$ frame.

The transformation from the frame $F$ to $F'$ is a 
clockwise rotation around the $y$-axis by an angle 
$\zeta$,
which is determined by
\begin{equation}
	\cos\zeta
	=
	\frac{(\textbf{p}_1+\textbf{k})\cdot\textbf{k}}{|\textbf{p}_1+\textbf{k}||\textbf{k}|}
	=
	\frac{|\textbf{p}_1|\cos\theta+|\textbf{k}|}{|\textbf{p}_1+\textbf{k}|}, 
    \quad 
    	\sin\zeta=
	\frac{|\textbf{p}_1|\sin\theta}{|\textbf{p}_1+\textbf{k}|}.
\end{equation}
In this case,
$|\textbf{p}_a|_y=|\textbf{p}_a'|_{y'}$,
and
\begin{align}
	&|\textbf{p}_a|_x
	=
	|\textbf{p}_a'|_{x'}\cos\zeta+|\textbf{p}_a'|_{z'}\sin\zeta=
	|\textbf{p}_a|
	(\sin\xi\cos\phi\cos\zeta+\cos\xi\sin\zeta),
	\\
	&|\textbf{p}_a|_z
	=
	|\textbf{p}_a'|_{z'}\cos\zeta
	-|\textbf{p}_a'|_{x'}\sin\zeta
	=|\textbf{p}_a|
	(\cos\xi\cos\zeta-\sin\xi\cos\phi\sin\zeta).
\end{align}
The three Mandelstam variables
are then 
given by
\begin{align}
&s
=(P_1+K)^2
=
m_\ell^2
+2|\textbf{k}|(E_1-|\textbf{p}_1|\cos\theta),
\label{eq:Mandelstam:s}
\\
&t
=(K-P_a)^2=m_a^2+2|\textbf{k}|
\left[
|\textbf{p}_a|
(\cos\xi\cos\zeta-\sin\xi\cos\phi\sin\zeta)
-E_a
\right],
\\
&u=m_\ell^2+m_{\ell'}^2+m_a^2-s-t.
\label{eq:Mandelstam:u}
\end{align}

\normalem
\bibliography{ref.bib}{}
\bibliographystyle{utphys28mod}

\end{document}